\documentclass{jfm}
\usepackage{graphicx}
\usepackage{epstopdf, epsfig}
\usepackage{textcomp}
\usepackage{amsmath}
\usepackage{xcolor}

\shorttitle{Streaming controlled by meniscus shape}
\shortauthor{Y. Huang et al.}

\title{Streaming controlled by meniscus shape}

\author{Y. Huang\aff{1}\aff{2}\footnote{yicheng.huang@stonybrook.edu} C.P. Wolfe\aff{2} J. Zhang\aff{3} J.-Q. Zhong\aff{1}\footnote{jinqiang@tongji.edu.cn}}

\affiliation{\aff{1}School of Physics Science and Engineering, Tongji University, Shanghai, China \aff{2}School of Marine and Atmospheric Sciences, Stony Brook University, Stony Brook, U.S. \aff{3}NYU-ECNU Institute of Physics at NYU Shanghai, Shanghai, China}

\begin{document}

	\maketitle

	\begin{abstract}
         Surface waves called meniscus waves often appear in the systems that are close to the capillary length scale.
         Since the meniscus shape determines the form of the meniscus waves, the resulting streaming circulation has a structure distinct from that caused by other capillary-gravity waves recently reported in the literature.
         In the present study, we produce symmetric and antisymmetric meniscus shapes by controlling boundary wettability and excite meniscus waves by oscillating the meniscus vertically.
         The symmetric and antisymmetric configurations produce different surface capillary-gravity wave modes and streaming flow structures. 
         The energy density of the streaming circulation increases at the rate of the fourth power of the forcing amplitude in both configurations.
         The flow symmetry of streaming circulation is retained under the symmetric meniscus, while it is lost under the antisymmetric meniscus.
         In our experiments, the streaming circulation primarily originates from the Stokes boundary layer beneath the meniscus and can be successfully explained using the existing streaming theory. 
	\end{abstract}


\section{Introduction}

Steady streaming is driven by Reynolds stresses and balanced by viscous stresses in the Stokes boundary layer \citep{batchelor2000introduction,riley2001steady}.
The momentum of the streaming flow in the boundary layer can be transferred into the liquid bulk through fluid viscosity to form an Eulerian mean circulation.
Streaming circulations play essential roles in fluid mass and passive scalar transport in many environments, such as inside blood veins, cochleae, and the boundary layer over ocean sediments \citep{lesser1972fluid, schneck1976pulsatile, holmedal2009wave}.
In recent years, streaming circulations driven by the surface waves have also been used to control the distribution of the suspended particles in the biomedical production and the food manufacturing \citep{chen2014microscale, strickland2015spatiotemporal, punzmann2014generation, francois2017wave}.
A more complete understanding of the underlying physical mechanisms that control the formation of streaming circulations is likely to be beneficial to these applications.

In this paper, we investigate the streaming flows inside a fluid system with an overlying free surface and a horizontal scale one order of magnitude larger than the capillary length, $l$.
When such a system is oscillated vertically, a variety of standing surface waves can be generated by the capillary effect \citep{lucassen1970properties,perlin2000capillary}.
The most well-known type of waves are Faraday waves \citep{faraday1831xvii,miles1990parametrically}, which are excited by parametric instability.
However, even when the system is stable to Faraday waves, one transverse surface wave is still observed if there is a static meniscus.
This \textit{meniscus wave} \citep{strickland2015spatiotemporal} is driven by periodic perturbations to hydrostatic balance due to the vertical oscillations \citep{antkowiak2007short}.
The resultant surface pressure anomaly deforms the liquid-air interface periodically, forcing a capillary-gravity wave \citep{douady1990experimental}.
Up to now, there is a lack of experiments which directly observe meniscus waves and their induced streaming circulation since meniscus waves are usually entangled with Faraday waves in most previous experiments \citep{douady1990experimental,strickland2015spatiotemporal}.

Streaming flows can occur along the free surface due to nonlinear coupling between surface deformations and the velocity field \citep{longuet1953mass,martin2002drift}.
In these systems, the geometry of the meniscus increases the complexity of the streaming flow.
Surfactants can promote the generation of streaming flows along the free surface by enhancing surface dissipation and tangential stress \citep{martin2006effect}.
In the fully contaminated limit, the streaming boundary layer generated by the free surface behaves in the same manner as that caused by a elastic solid surface \citep{perinet2017streaming, moisy2018counter}. 
How the static meniscus shape affects the streaming circulation structure in this limit remains an open question of great interest.

In the following sections, we present measurements of meniscus waves and the resultant streaming circulation under an oscillatory free surface.
These measurements are performed in a fluid system where the Faraday waves are inhibited.
By selecting the wettability of the lateral boundaries, we develop two kinds of meniscus with distinct symmetries.
When the lateral walls are both hydrophilic, a symmetric meniscus with two identical acute angles is created.
The resultant two-node linear meniscus wave gives rise to a four-eddy streaming circulation pattern.
Alternatively, the meniscus has an antisymmetric shape in a hydrophilic-hydrophobic wetting configuration.
In this case, the resultant linear meniscus wave has a single node and a double-eddy streaming circulation pattern is produced.
The energy of the streaming circulation scales like $e_{k}\sim(a_{0}/g_{0})^{4}$; the single power function indicating a single dynamical regime over the parameter range studied.
Remarkably, we further discover that for the two meniscus shapes we use, the left-right symmetry of the streaming flow pattern differs markedly under various forcing amplitudes $a_{0}/g_{0}$.
For the symmetric case, the streaming pattern remains highly symmetric in the full range of the applied forcing amplitude.
In the antisymmetric case, however, the streaming pattern becomes less symmetric as $a_{0}/g_0$ increases owing to the interaction between the static meniscus profile and the meniscus wave.
To explain these observations, we establish a simple analytical model to estimate the meniscus wave at linear order.
We then compute the streaming velocity distribution in the boundary layer following the method of \cite{gordillo2014measurement} and \cite{perinet2017streaming} and use this to drive a steady laminar flow model that reproduces the observed streaming velocity fields.

\section{Experiment}

\subsection{Experimental setup}

\begin{figure}
  \centerline{\includegraphics[scale=0.9]{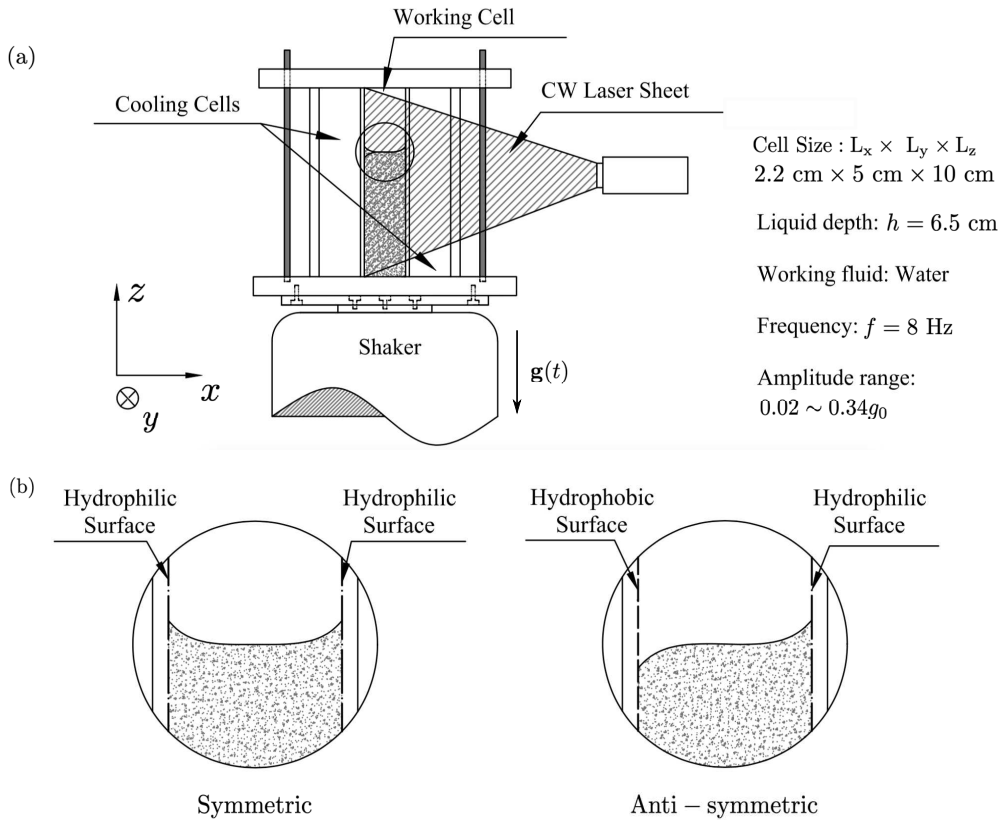}}
\caption{(a) Schematic drawing of the experimental apparatus.
A water tank was oscillated sinusoidally in the $z$ direction with a time-varying gravitational acceleration $g(t)=-(g_{0}+a_{0}\sin 2\pi ft)$. 
The tank was divided into three cells by two $\rm{SiO}_{2}$ glass plates.
The fluid velocity field was measured in the working cell using PIV.
A DPSS laser sheet illuminated the seeding particles in the working cell and the particle motion was recorded by a high-speed camera.
(b) Schematic of meniscus in symmetric and antisymmetric configurations. }
\label{fig:figure1}
\end{figure}

In the experiment, we used a rectangular Plexiglas tank divided into three cells by two optical glass plates (figure \ref{fig:figure1}a).
The middle cell was the working cell, with the dimensions $L_{x}=2.2~\rm{cm}$ in length, $L_{y}=5.0~\rm{cm}$ in width and $L_{z}=10.0~\rm{cm}$ in height.
Deionized water filled the working cell to a depth of $h=6.5~\rm{cm}$.
Two types of meniscus symmetry were created by selecting the wettability conditions of the glass plates (see figure \ref{fig:figure1}b).
The clean optical glass surface was hydrophilic with a contact angle close to $45^{\circ}$.
A hydrophobic boundary was made by applying a coating of hydrophobic nano $\rm{SiO_{2}}$ film (CHEMNANO\textsuperscript{\textregistered} NC317) to an optical glass surface, creating a contact angle close to $135^{\circ}$.
A working cell with two hydrophilic lateral boundaries creates the symmetric setting while choosing one boundary hydrophobic and the other hydrophilic creates an antisymmetric system.
The contact angle of water with the rest Plexiglas boundary was close to $90^{\circ}$.

The fluid tank was installed on an electrodynamic shaker (Labworks\textsuperscript{\textregistered} ET-139) that vibrated sinusoidally in the $z$ direction with an excess gravitation acceleration varying from $0.02g_{0}\sim 0.34 g_{0}$ and a frequency of $f=8~\rm{Hz}$.
The control signal produced from a wave-function generator was amplified using a linear power amplifier to drive the shaker.
An accelerometer mounted on the base plate of the fluid tank (not shown in figure \ref{fig:figure1}) measured the instantaneous acceleration that was fed back to the wave-function generator to regulate the control signal.
The closed-loop control system reduced error and ensured the stability of the shaker oscillation.

We measured the fluid velocity field in the working cell using a Particle Imaging Velocimetry (PIV) system.
Neutrally-buoyant particles of diameter $10~\rm{\mu m}$ were seeded in the flow.
Because of the effect of surface tension, some of the seeding particles were suspended on the free surface.
We used these particles to track the surface deformation (see section 2.2).
We found that these surface particles also played a role in enhancing surface damping and surface-deformation induced streaming.
The PIV particles in the system was illuminated from the right using a continuous, diode pump solid-state (DPSS) laser, creating a $1~\rm{mm}$ thick light-sheet that passed through a vertical cross-section and the mid-plane in the $y$ direction of the cell. 
PIV images that covered a region from the surface to a fluid depth of around $3~\rm{cm}$ were captured by high-speed camera at a frame rate of $80~\rm{Hz}$, achieving a time resolution of 10 frames per oscillation period.
Two-dimensional velocity maps were obtained by cross-correlating either two consecutive images (for the primary flow) or two images with the same phase in two consecutive periods (for the secondary flow).  
Each velocity vector was calculated from interrogation windows spanning $32{\times}32$ pixels, with 50$\%$ overlap of neighbouring sub-windows.
Each vector covers a region of $16{\times}16$ pixels and there are $80{\times}64$ velocity vectors per frame, achieving a spatial resolution of $0.34~\rm{mm}$ (one order in magnitude smaller than the capillary length scale). 

The laser light sheet may heat up the fluid on the right side of the working cell and give rise to an overturning circulation.
The speed of this overturning flow could be up to the order of $10^{-3}~\rm{cm/s}$, comparable with the speed of the streaming flow we study.
To avoid this laser-induced convection, the two lateral cells were filled with water to reduce the thermal heating of the fluid in the working cell.
With this thermal protection, the speed of the overturning flow was reduced to $\mathcal{O}\left(10^{-5}~\rm{cm/s}\right)$ and was negligible.

\subsection{Observation of the surface deformation}

\begin{figure}
  \centerline{\includegraphics[scale=0.84]{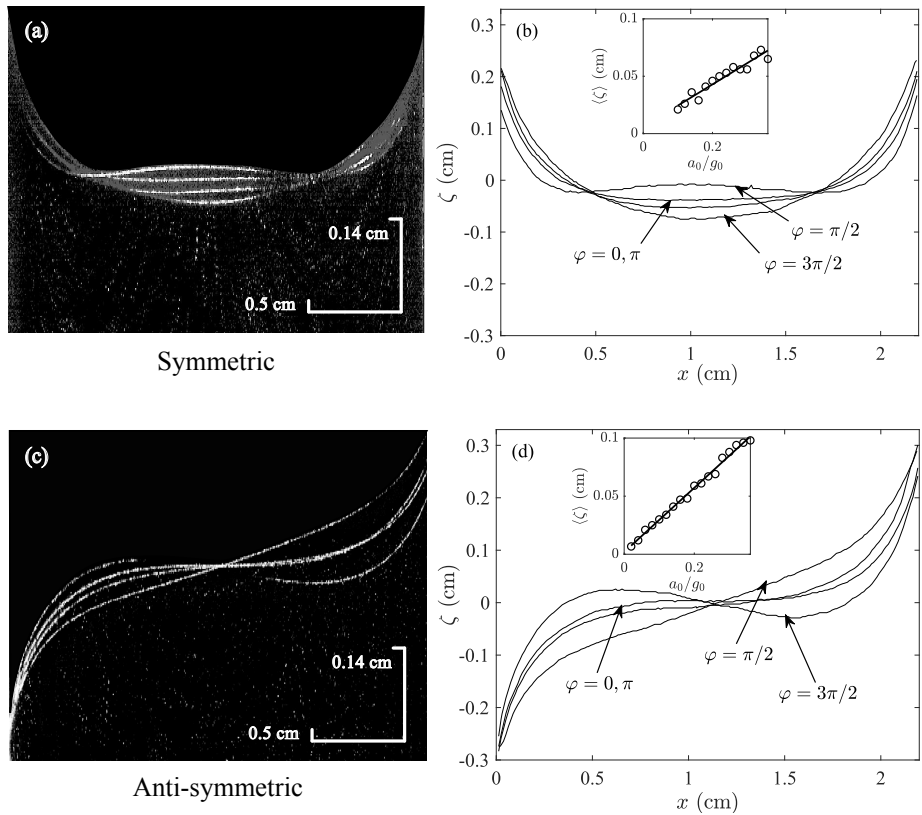}}
  \caption{Images and surface profiles of meniscus shape deformation at $\phi=0,\pi/2, \pi, 3\pi/2$ with $f=8~\rm{Hz}$ and $a_{0}=0.2g_{0}$.
  (a) PIV image of the vibrating surface in the symmetric configuration.
  (b) The extracted surface profiles corresponding to (a).
  (c) PIV image of the vibrating surface in the antisymmetric configuration.
  (d) The extracted surface profiles of symmetric scenario corresponding to (c).
  The insets in (b) and (d) shows the horizontally averaged surface deformation amplitude $\langle \zeta\rangle$ as a function of $a_{0}/g_{0}$.
  The symbols in the insets are experimental measurements and the solid line is a linear regression.
}
\label{fig:figure2}
\end{figure}

The time varying acceleration of the fluid tank in the lab frame is $a(t) = a_0 \sin\varphi(t)$, where $a_{0}$ is the forcing amplitude and $\varphi(t)=2\pi ft$ is the oscillation phase. From Newton's third law, the effective gravity felt by the fluid in the tank is 
\[ 
	g(t)=-\left[g_{0}+a_{0}\sin\varphi(t)\right],
\]
in which $g_{0}$ is the gravitational acceleration constant.
When $\varphi=\pi/2$, the fluid tank is at its lowest position with maximum super-gravity; when $\varphi=3\pi/2$, the tank is at its highest position with maximum sub-gravity.
To force the system without exciting the Faraday instability, we select $f=8.0~\rm{Hz}$ and $0.02\le a_0/g_0 \le 0.34$ (see Appendix A).
Thus the periodic deformation of the liquid-air interface resonates with the cavity modes, giving rise to standing meniscus waves.
We observe that such a surface deformation mode is strongly dependent on the geometry of the static meniscus.
To quantify the relation between this surface deformation mode and the shape of the static meniscus, we measure the surface profile in different oscillation phases from the PIV images.
It is found that a large proportion of the liquid-air interface is clearly visible due to seeding particles accumulated at the interface which are illuminated by the laser sheet (figure \ref{fig:figure2}a and \ref{fig:figure2}c).
From these images we extract the surface profiles $\zeta(x)$ at four oscillatory phases $\phi=0$, $\pi/2$, $\pi$ and $3\pi/2$ as shown in figure \ref{fig:figure2}b and \ref{fig:figure2}d.

The meniscus oscillates at the frequency $f$.
The response of meniscus wave amplitude $\zeta(x)$ to the forcing acceleration amplitude $a_0/g_0$ is estimated by the horizontally averaged surface deformation amplitude  $\langle \zeta\rangle$, which is defined by
\[
               \langle \zeta\rangle=\overline{|\zeta_{\varphi=\pi/2}-\zeta_{\varphi=3\pi/2}|},
\]
where $\zeta_{\varphi=\pi/2}$ and $\zeta_{\varphi=3\pi/2}$ are the surface profiles captured at $\varphi=\pi/2$ and $\varphi=3\pi/2$ respectively; the overline denotes an average on $x$ direction.
The mean amplitude of the meniscus oscillation $\langle\zeta\rangle$ grows linearly with the forcing amplitude (see the insets figure \ref{fig:figure2}b, \ref{fig:figure2}d).
In the symmetric case, the excited mode is symmetric and has two nodes.
At maximum super-gravity ($\varphi=\pi/2$) the liquid surface is bulging from the middle of the cell, analogous to the phenomenon of liquid jet observed before \citep{antkowiak2007short}.
At maximum sub-gravity ($\varphi=3\pi/2$), the meniscus bends downwards to its greatest extent.
Between these two states, the fluid flows periodically between the lateral boundary and the middle of the cell.
In the antisymmetric case, an oscillatory mode with one node at the center ($x=L/2$) is instead excited.
When the system is at maximum sub-gravity ($\varphi=3\pi/2$), the fluid surface forms an '$S$' shape, with the convex part on the hydrophobic side ($x<L/2$) and the concave part on the hydrophilic side ($x>L/2$).
When the system is at maximum super-gravity ($\varphi=\pi/2$), the fluid surface approaches a straight slope.
Between these two states, the fluid sloshes between the two lateral boundaries.
No observable surface oscillation is found on the $y$ direction.
Representing the mode number as $(l,m)$, where $l$ and $m$ are positive integers denoting the mode number on $x$ and $y$ axes respectively, the symmetric and antisymmetric configurations give rise to modes $(2,0)$ and $(1,0)$ waves, respectively.
The PIV results indicated that the oscillatory motion decays exponentially with depth, so these are deep water waves.

\subsection{Observation of the streaming motion}

The streaming circulation is observed when the PIV images are broadcast stroboscopically.
The streaming circulation is one order of magnitude weaker than the instantaneous primary flow.
Once the circulation forms, it persists as long as the forcing oscillation continues.
During the experiment, we collected the PIV images one minute after the forcing was turned on, when a steady streaming circulation had been established.
The flow velocity and the vertical scale of the circulations, for both wettability configurations, appear to increase if the oscillation amplitude $a_{0}$ increases.
The streaming circulation has distinct structures in symmetric and antisymmetric cases.

\begin{figure}
  \centering
  \includegraphics[scale=0.9]{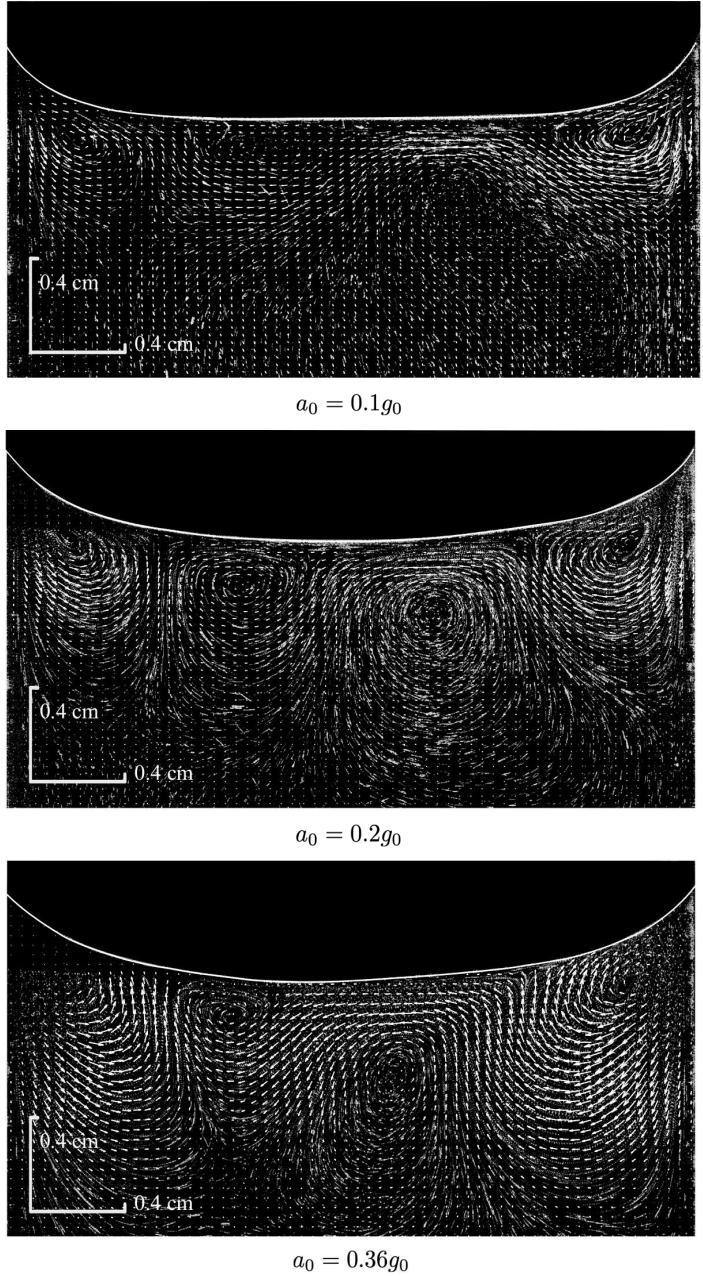}
  \caption{Velocity fields of the streaming flows overlying the raw PIV image in the symmetric case for three values of $a_0$.
  For each case, the driving frequency is $f=8~\rm{Hz}$.
  The liquid-air interface has been highlighted manually.}
\label{fig:figure4}
\end{figure}

In the symmetric case (figure \ref{fig:figure4}), the secondary circulation has a stable four-vortex structure, with clockwise (counterclockwise) rotation near the right (left) lateral boundary.
When $a_{0}\sim0.1g_{0}$, the central vortices are not clearly visible in the PIV image.
As the forcing gets stronger, the central vortices become apparent with approximately circular streamlines.
At the strongest forcing amplitude, the boundary vortices are well developed, whereas the central vortices become suppressed and asymmetric.

\begin{figure}
  \centerline{\includegraphics[scale=0.9]{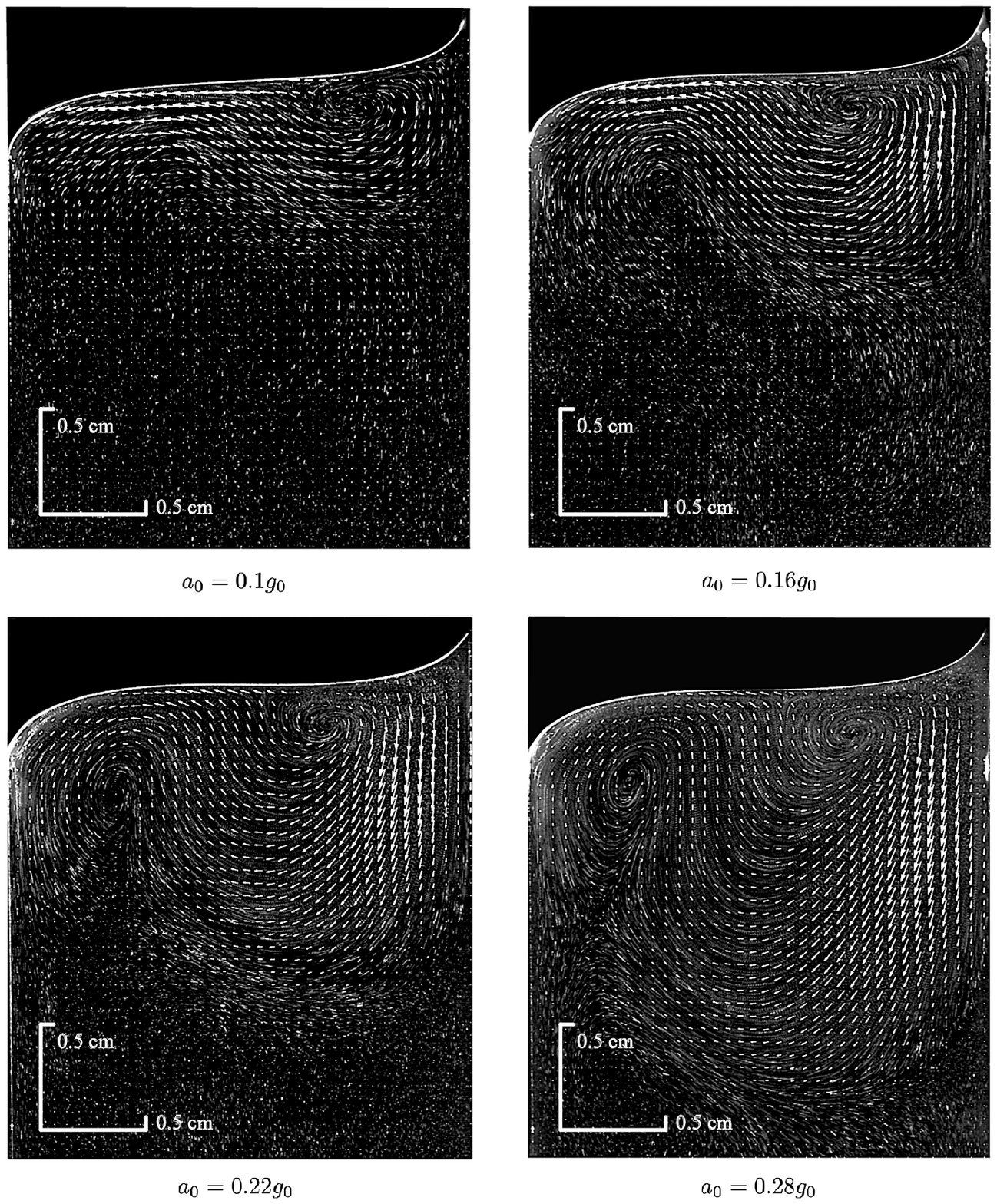}}
  \caption{Velocity fields of the streaming flows overlying the raw PIV image in the antisymmetric case for four values of $a_0$.
  For each case, the driving frequency is $f=8~\rm{Hz}$.
  The liquid-air interface has been highlighted manually.}
\label{fig:figure3}
\end{figure}

In the antisymmetric case, the secondary flow field has a dipole circulation structure with a weaker counterclockwise circulation near the hydrophobic side and a stronger clockwise circulation near the hydrophilic side (figure \ref{fig:figure3}).
When the forcing is weak, $a_{0}\sim0.1g_{0}$, the counterclockwise circulation is only partially visible in the PIV image, but a strong leftward velocity is found below the free surface.
As the forcing amplitude increases, the counterclockwise circulation becomes more dominant and downward motion becomes more prominent near the hydrophobic side.

\begin{figure}
  \centerline{\includegraphics[scale=0.46]{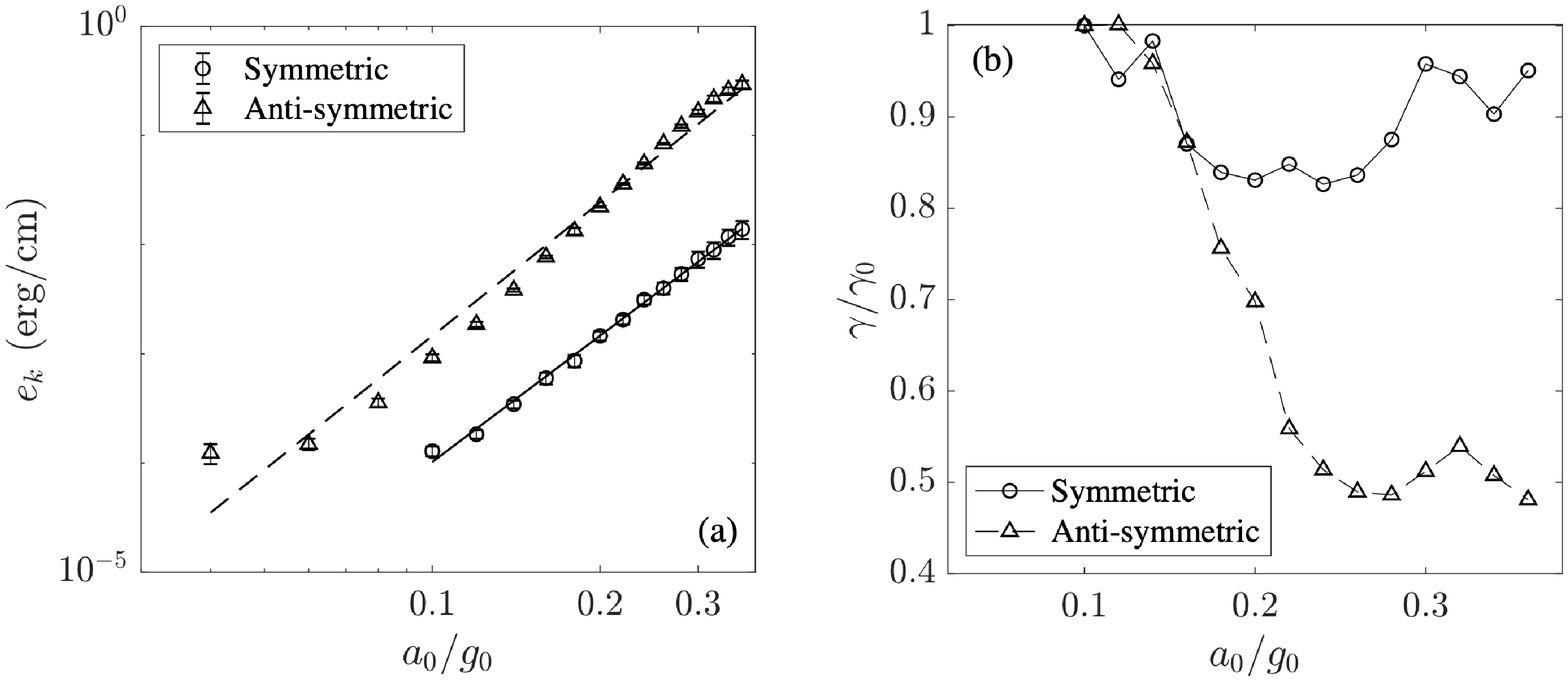}}
  \caption{Response of the system to changes in forcing amplitude $a_{0}/g_{0}$.
$a_{0}/g_{0}$ is increased from $0.1$ to $0.34$ for symmetric case; from $0.04$ to $0.34$ for antisymmetric case from experiment.
  (a) The response of energy density $e_{k}$.
  The fitted trends give $e_k{\sim}(a_{0}/g_{0})^4$ and $e_k{\sim}(a_{0}/g_{0})^{3.9}$ for the symmetric and antisymmetric cases, respectively.
  (b) The response of symmetry parameter $\gamma$, normalized by $\gamma_{0}=\gamma|_{a_{0}/g_{0}=0.1}$.}
\label{fig:figure5}
\end{figure}

To quantify the dynamical behavior of the secondary flow, we integrate the streaming flow velocity on the observed section to determine the kinetic energy density at $y$-direction:
\begin{eqnarray}
	e_{k}=\frac{1}{2}\rho\iint_{S}(u^{2}+v^{2})\,dx\, dz.
\label{energy_integral}
\end{eqnarray}
As shown in figure \ref{fig:figure5}a, the kinetic energy density for fully-established secondary flow structure $e_{k}$ scales like 
\begin{eqnarray}
	e_{k}\sim (a_{0}/g_{0})^{\beta},
\label{scaling_law}
\end{eqnarray}
where $\beta \approx 4$ for both symmetric and antisymmetric cases.
This indicates that the secondary circulation is caused by the quadratic nonlinearity of the velocity field, which is consistent with the direct numerical simulations of \cite{carrion2017mean}.
Note that the flow velocity near the boundary is not resolvable by PIV, so the integration in \eqref{energy_integral} excludes the region within $0.034$ cm of the lateral boundaries. 
Further, the camera does not capture the whole water bulk, so the integration covers the region $z > -1.79~\rm{cm}$ in the symmetric case and  $z > -1.65~\rm{cm}$ in the antisymmetric case.

The spatial symmetry of the secondary circulation pattern is quantified by the symmetry parameter
\begin{align}
	\gamma=\frac{\iint_{x<L/2}dx\,dz |\Omega(x,z)|/A_{x<L/2}}{\iint_{x>L/2}dx\,dz |\Omega(x,z)|/A_{x>L/2}}.
\label{sd}
\end{align}
which denotes the ratio of mean vorticity magnitude of the left region over that of the right region of the working cell.
In equation \eqref{sd}, $|\Omega(x,z)|$ is the magnitude of the vorticity, $A_{x>L/2}$ and $A_{x<L/2}$ are the liquid bulk area at the right or left side, respectively, of $x=1.1~\rm{cm}$ below the static meniscus.
The right panel of figure \ref{fig:figure5} shows the variation of $\gamma/\gamma_{0}$ against $a_{0}/g_{0}$, where $\gamma_{0}=\gamma|_{a_{0}/g_{0}=0.1}$ is a reference $\gamma$ value.
The variation of $\gamma/\gamma_{0}$ with forcing amplitude is different in the symmetric and antisymmetric cases.
In the symmetric case, the secondary flows remains highly symmetric, with no obvious trend in $\gamma/\gamma_{0}$.
In contrast, in the antisymmetric case the symmetry parameter $\gamma$ decreases rapidly as the forcing amplitude increases then saturates at $\gamma \approx 0.5\gamma_0$ for large forcing amplitudes.

\section{Theory} \label{sec:theory}

In this section, we develop a theoretical description of the meniscus wave and resulting secondary circulation. 
We start by deriving the static shape of the meniscus and solve for the wave modes on this background state in section~\ref{sec:theory:wave}. 
The secondary circulation is obtained in section~\ref{sec:theory:streaming} by determining the streaming circulation along the free surface and lateral boundaries and using these as boundary conditions for a steady hydrodynamics solver. 

\subsection{Meniscus wave} \label{sec:theory:wave}

The meniscus wave is solved by linearization of the ideal fluid equations with a free surface \citep[e.g.,][]{batchelor2000introduction}. 
The nonlinear fluid equations are
\begin{align}
	\nabla^{2}\phi &= 0,\label{eqn:GE1}\\
	\partial_{t}\phi|_{z=\zeta}+[g_{0}+a_{0}\sin(\omega t)]\zeta &=\frac{\sigma}{\rho}\frac{\partial_{xx}^{2}\zeta}{[1+(\partial_{x}\zeta)^{2}]^{3/2}},\label{eqn:GEBC1}\\
	\partial_{t}\zeta+\partial_{x}\phi|_{z=\zeta}\partial_{x}\zeta &= \partial_{z}\phi|_{z=\zeta},\label{eqn:GEBC2}\\
	\phi|_{z\rightarrow-\infty}&=0,\label{eqn:GEBC3}\\
	\partial_{x}\phi|_{x=0,L}&=0,\label{eqn:GEBC4}
\end{align}
where $\zeta=\zeta(x,t)$ is the surface elevation and $\phi=\phi(x,z,t)$ is the velocity potential field.
The parameter notation is the same as in the previous section. 
The cell width is $L=2.2~\rm{cm}$, $g_{0}=980~\rm{cm/s^{2}}$ is the gravitational acceleration, $\rho=1~\rm{g/cm^{3}}$ is the density of water, $\sigma=72~\rm{dyne/cm}$ is the surface tension of water, $a_{0}$ is the forcing acceleration, and $\omega=2\pi f$ is the forcing angular frequency.

The linearization is done in the following way:
We use the capillary length $l=\sqrt{\sigma/(\rho g_{0})} = 0.27~\rm{cm}$ as the characteristic length scale in the $x$- and $z$-directions.
The scale of the surface displacement due to the forced oscillation is $a_{0}/\omega^2\approx 0.008\textendash0.13~\rm{cm}$, so the ratio between these two length scales is a small dimensionless number, $\epsilon=a_{0}/(l\omega^2)\approx0.03\textendash0.5$ that scales the linear system response.
In the experiment, the dimensionless forcing strength $a_{0}/g_{0}$ has the magnitude range of $0.02\textendash0.32$, which is $O(\epsilon)$.
After scaling the velocity potential, $\phi$, by $g_{0}a_{0}/\omega^{3}$, the solution of the ideal fluid equations \eqref{eqn:GE1}--\eqref{eqn:GEBC4} can be expanded in an asymptotic series in powers of $\epsilon$:
\begin{align}
\zeta&=\zeta_{0}+\zeta_{1},\label{eqn:AsyExp_1}\\
\phi&=\phi_{0}+\phi_{1},\label{eqn:AsyExp_2}
\end{align}
The quantities with the subscript $0$ are $O(1)$, which represents a situation when the forcing is neglected and depicts the static balance in the fluid bulk.
Thus, both $\phi_{0}$ and $\zeta_{0}$ are independent of time.
The quantities with subscript $1$---that is, $O(\epsilon)$---represent the linear wave forced by the vertical oscillation.
As the motion decays exponentially from the free surface and the depth is $3$ times larger than cell width, we apply an infinite-depth bottom boundary condition.

\subsubsection{Zeroth order solution} \label{sec:theory:wave:zeroth}

\begin{figure}
  \centerline{\includegraphics[scale=0.46]{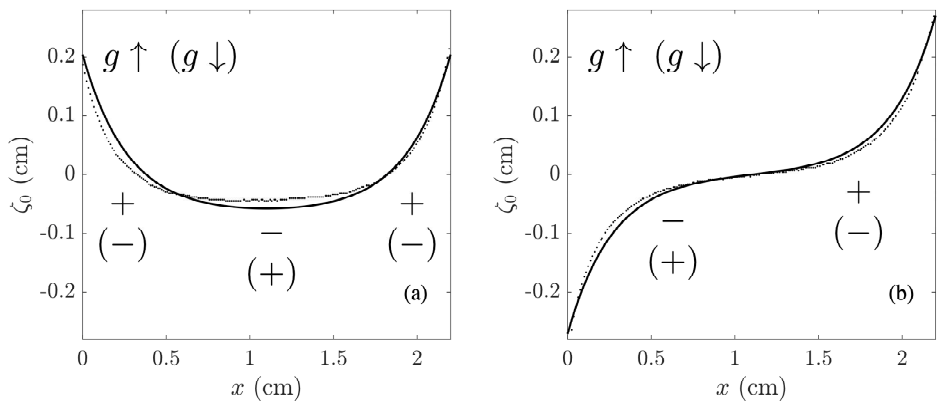}}
  \caption{The zeroth-order meniscus solution $\zeta_{0}$ (solid line) compared to the experimental measurement (dotted line) for the (a) symmetric and (b) antisymmetric cases.
  The $+$ and $-$ signs represent the pressure surplus and deficit, respectively, that arises when the acceleration is changed. 
  The pressure anomaly distribution out of (in) the bracket is for the situation when $g$ increases (decreases).}
\label{fig:figure6}
\end{figure}

At $O(1)$, \eqref{eqn:GE1}--\eqref{eqn:GEBC1} become the Young-Laplace equation:
\begin{eqnarray}
\partial_{xx}^{2}\zeta_{0}=\frac{\rho g_{0}}{\sigma}(\zeta_{0}-z^{*}).\label{eqn:ZOE}
\end{eqnarray}
Since the left plate and right plate completely prevent the liquid mass exchange, the experiment is done with total fluid volume conserved. 
The reference elevation, 
\begin{eqnarray}
	z^{*}=-\frac{\sigma}{\rho g_{0} L}(\partial_{x}\zeta_{0}|_{x=L}-\partial_{x}\zeta_{0}|_{x=0})=-\frac{\sigma}{\rho g_{0}L}[\tan(\theta_{r})-\tan(\pi/2-\theta_{l})].
\end{eqnarray}  
is determined by integrating equation \eqref{eqn:ZOE} in $x$, applying the lateral boundary conditions, and requiring that the horizontal integral of $\zeta_0$ vanish.
The quantities $\theta_{l}$ and $\theta_{r}$ is the contact angles at the left and right boundaries, respectively.
Following the experimental setup, in the symmetric case $\theta_{l}=\theta_{r}=45^{\circ}$ and $\theta_{l}=135^{\circ}$ and $\theta_{r}=45^{\circ}$ in the antisymmetric case.
The corresponding boundary conditions for equation (\ref{eqn:ZOE}) are written as
\begin{align}
	\partial_{x}\zeta_{0}|_{x=0}&=-s,\\
	\partial_{x}\zeta_{0}|_{x=L}&=1.
\end{align}
Here, $s$ controls the symmetry of the meniscus; it is $1$ in the symmetric case and $-1$ in the antisymmetric case.
These boundary conditions and the above zeroth-order equation are satisfied by
\begin{eqnarray}
\zeta_{0}=\frac{(1+s\cosh(mL))}{m\sinh (mL)}\cosh(mx)-\frac{s}{m}\sinh(mx)+z^{*},
\end{eqnarray}
where $m=l^{-1}$.
In figure (\ref{fig:figure6}), the solutions are compared to the static surface profile measurements for both symmetric and antisymmetric cases---the theoretical result captures the shape of the free surface well.

\subsubsection{First order solution} \label{sec:theory:wave:first}

According to the surface pressure balance condition \eqref{eqn:GEBC1}, a change in the gravitational acceleration $g$ results in horizontally varying pressure differences along the meniscus.
Taking the mean surface elevation to be $z=0$, a rising of gravitational acceleration produces a positive pressure difference in the region where $\zeta>0$ and a negative pressure difference where $\zeta<0$.
The pressure difference distribution is inverted for falling gravitational acceleration.
The distribution of pressure anomalies illustrated in figure \ref{fig:figure6} lead to a wave with two nodes [a $(2,0)$ wave] in the symmetric case and a wave with a single node [a $(1,0)$ wave] in the antisymmetric case.

\begin{figure}
  \centerline{\includegraphics[scale=0.46]{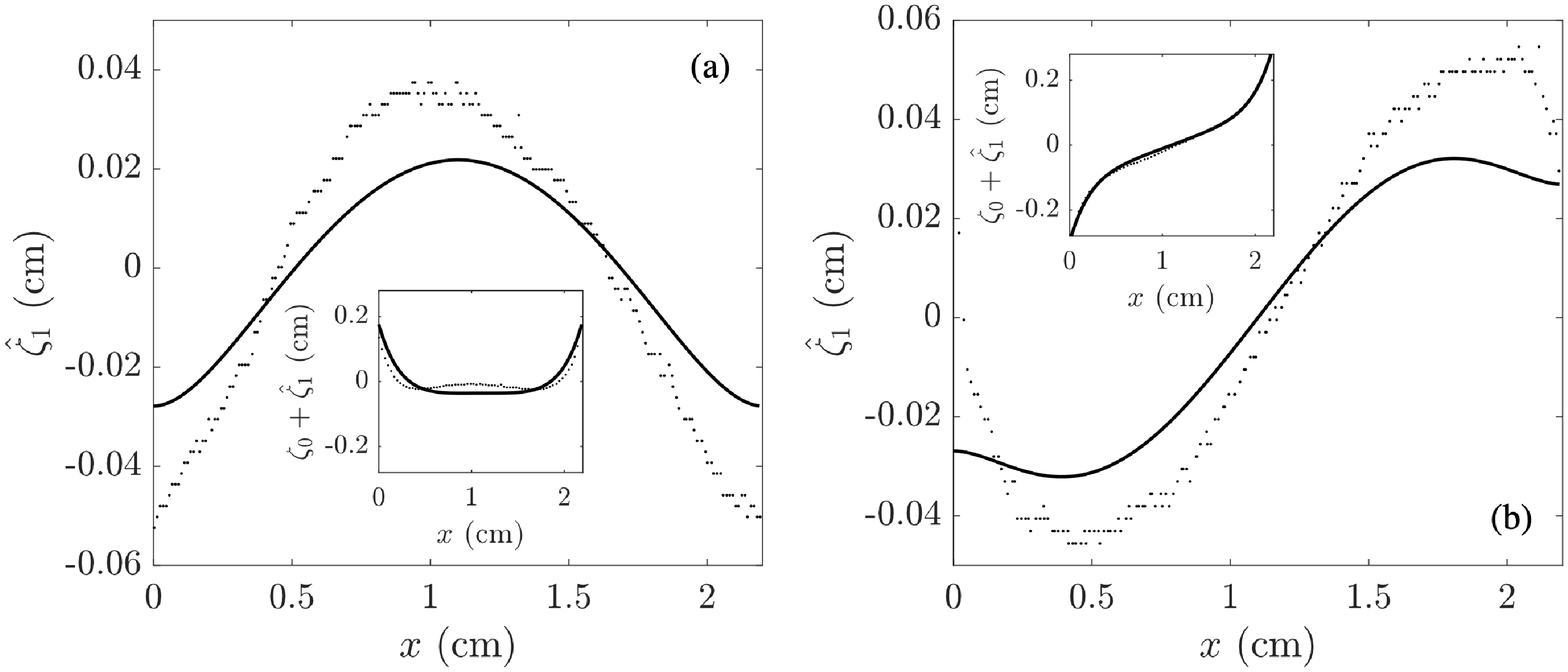}}
  \caption{The first order meniscus solutions $\hat{\zeta}_{1}$ (solid line) compared to the experimental measurements (dotted line) when $a_{0}=0.2g_{0}$ and $f=8~\rm{Hz}$ for the (a) symmetric and (b) antisymmetric cases.
  The experimental $\hat{\zeta}_{1}$ are produced by subtracting the static surface profiles from the oscillating surface profiles captured during the slack phase.
  The insets show the combination of $\zeta_{0}$ and $\hat{\zeta}_{1}$ solutions and the correspondent experiment measurements. }
\label{fig:figure7}
\end{figure}

At $O(\epsilon)$, the governing equations and boundary conditions become
\begin{align}
	\nabla^{2}\phi_{1}&=0,\label{eqn:FOE1}\\
	\partial_{t}\phi_{1}|_{z=0}+g_{0}\zeta_{1}-\frac{\sigma}{\rho}\partial_{xx}^{2}\zeta_{1}&=-a_{0}\zeta_{0}\sin(\omega t),\label{eqn:FOEBC1}\\
	\partial_{z}\phi_{1}|_{z=0}-\partial_{t}\zeta_{1}&=0,\label{eqn:FOEBC2}\\
	\phi_{1}|_{z=-\infty}&=0,\label{eqn:FOEBC3}\\
	\partial_{x}\phi_{1}|_{x=0,L}&=0.\label{eqn:FOEBC4}
\end{align}
The solution that satisfies \eqref{eqn:FOE1}--\eqref{eqn:FOEBC4} takes the form
\begin{align}
\label{Solution_phi}
	\phi_{1}&=\hat{\phi}_{1}\cos(\omega t)=\sum\limits_{n=1}^{\infty}A_{n}e^{k_{n}z}\cos(k_{n}x)\cos(\omega t),\\
\label{Solution_zeta}
	\zeta_{1}&=\hat{\zeta}_{1}\sin(\omega t)=\sum\limits_{n=1}^{\infty}\frac{A_{n}k_{n}}{\omega}\cos(k_{n}x)\sin(\omega t),
\end{align}
where the eigen-wavenumber $k_{n}=n\pi/L$ and coefficients $A_{n}$ are given by
\begin{align}
	A_{n}&=\frac{2\omega a_{0}}{\omega^{2}-\omega_n^2}\frac{(-1)^{n}+s}{(m^{2}+k_{n}^{2})L},
\intertext{where}
	\omega_n^2 &= g_0 k_n + \frac{\sigma}{\rho}k_n^3
\end{align}
is the (squared) natural frequency of a gravity-capillary wave with wavenumber $k_n$.
Note that these wave modes [\eqref{Solution_phi} and \eqref{Solution_zeta}] are not parametric resonance modes (Faraday modes), but ordinary resonance modes since they have the same frequency as the forcing.
The wave energy is concentrated around those $n$ for which $\omega_n$ is close to $\omega$.
If $\omega = \omega_n$, for a given $n$, the capillary-gravity wave can travel integer multiples of the cell length during one forcing cycle.
When $\omega=16\pi~\rm{rad}^{-1}$ as in our experiment, both modes $n=1$ and $n=2$ are close to resonance.
In our case, the basin modes are further selected by the symmetry of the resting meniscus, since $A_n = 0$ for $n$ odd in the symmetric case and $A_n = 0$ for $n$ even in the antisymmetric case.
Thus, only the $n = 2$ ($n = 1$) mode is close to resonance in the symmetric (antisymmetric) case.
The amplitude $A_{n}$ decays like $O(n^{-5})$, so the series converges quickly.
In figure \ref{fig:figure7}, the infinite series is terminated at $n = 200$ to produce the estimate of the meniscus wave amplitude $\hat{\zeta}_{1}$ for both symmetric and antisymmetric cases.

From the comparison in figure \ref{fig:figure7}, we can see that our theory captures about $50\%$ variance of the experiment value at first order.
The linearized theory is inadequate to produce a more accurate result, because our basic scaling approximation $\partial_{x}(\zeta_{0}+\zeta_{1})\rightarrow0$ is not satisfied near the contact angle.
In our system, $\partial_{x}\zeta_{0}=1$, obviously violates the approximation.
To estimate the magnitude of $\partial_{x}\zeta_{1}$, we use the contact line boundary condition from \cite{hocking1987waves}
\begin{eqnarray}
    \partial_{t}\zeta=c_{s}(\partial_{x}\zeta-\partial_{x}\zeta_{0}). \label{contact_line}
\end{eqnarray}
Here, $c_{s}$ is the slipping parameter of contact line.
Under this new relationship, the gradient of first order surface elevation $\partial_{x}\zeta_{1}$ can be scaled by $a_{0}/(\omega c_{s})$.
According to the experiment measurement done by \cite{cocciaro1993experimental}, the slipping parameter $c_{s}=D l a_{0}$, in which $D=0.5\pm0.05~\rm{s/cm}$ is a fitting constant (it is necessary to note that in their experiment the working fluid is water, solid surface is Plexiglas and the contact angle is at $62^{\circ}$).
$\partial_{x}\zeta_{1}\sim 0.15$ in our experiment is a small number.
So, the solution we give in this section can be valid near the contact line only when contact angles are both close to $90^{\circ}$, when $\partial_{x}\zeta_{0}$ is close to 0.
Nevertheless, to get a precise analytical model of meniscus wave is not the purpose of this work.
In the next section, we will show that such a meniscus wave solution is sufficient to explain the streaming circulation pattern formation.

\subsection{Streaming circulation} \label{sec:theory:streaming}

In this section, we explain the secondary circulation pattern shown in figures~\ref{fig:figure4} and \ref{fig:figure3}.
Firstly, we note that as the dominant mode number of primary flow doubles, the number of secondary vortices doubles as well.
Moreover, the nodes of the meniscus wave are above upwelling flow, while the anti-nodes cap downwelling flows.
The spatial relationships between the meniscus wave and the streaming flow indicates that the streaming flow is generated by the meniscus wave itself.
There are a large number of seeding particles captured by the free surface, forming a membrane (see figure~\ref{fig:figure2}). 
\cite{moisy2018counter} has shown that under these conditions, the free surface acts like a no-slip boundary.
We thus hypothesize that such conditions generate a Stokes boundary layer at the free surface which drives a secondary streaming flow in the boundary layer.
This streaming boundary layer affects the bulk through viscous momentum transfer, so as to generate the secondary circulation pattern.

\subsubsection{Streaming under the free surface}

Following \cite{perinet2017streaming}, the streaming velocity under a meniscus fully contaminated by seeding particles can be written as
\begin{align}
	{\mathbf{u}}_{s}|_{z=\zeta_{0}}=-\frac{3}{4\omega}\hat{u}_{\parallel}\partial_{\tau}\hat{u}_{\parallel}\hat{{\bf{t}}}
	\label{streaming_velocity_1}
\end{align}
where $u_{||}$ denotes the primary flow parallel to the meniscus, $\hat{{\bf{t}}}$ is the tangential unit vector along the meniscus.
The tangential velocity at the meniscus, $u_{\parallel}$, is obtained by linear extrapolation from $z = 0$ (see Appendix B).
By substituting the expression for $u_{\parallel}$ into \ref{streaming_velocity_1}, we find
\begin{multline}
	{\mathbf{u}}_{s}|_{z=\zeta_{0}}= \Bigg\{-\frac{3}{4\omega}\sum_{n,m}A_{n}A_{m}(k_{n}+\zeta_{0}k_{n}^{2})(k_{m}+\zeta_{0}k_{m}^{2})\\
	\left[k_{m}-\frac{\rho g_{0}}{\sigma}(\zeta_{0}-z^{*})\right]\sin(k_{n}x)\cos(k_{m}x)+O(\partial_{x}\zeta_{0})\Bigg\}\hat{{\textbf{\i}}}+O(\partial_{x}\zeta_{0})\hat{{\textbf{\j}}},
	\label{streaming_velocity_2}
\end{multline}
where $\hat{\textbf{\i}}$ and $\hat{\textbf{\j}}$ are the horizontal and vertical unit vectors, respectively.
From this result, we can see that the surface profile $\zeta_{0}$ modifies the streaming velocity distribution.
To be consistent with the scaling assumptions of the primary flow solution, we have neglected terms that are $O(\partial_{x}\zeta_{0})$, which eliminates the vertical velocity along the meniscus.

\begin{figure}
  \centerline{\includegraphics[scale=0.46]{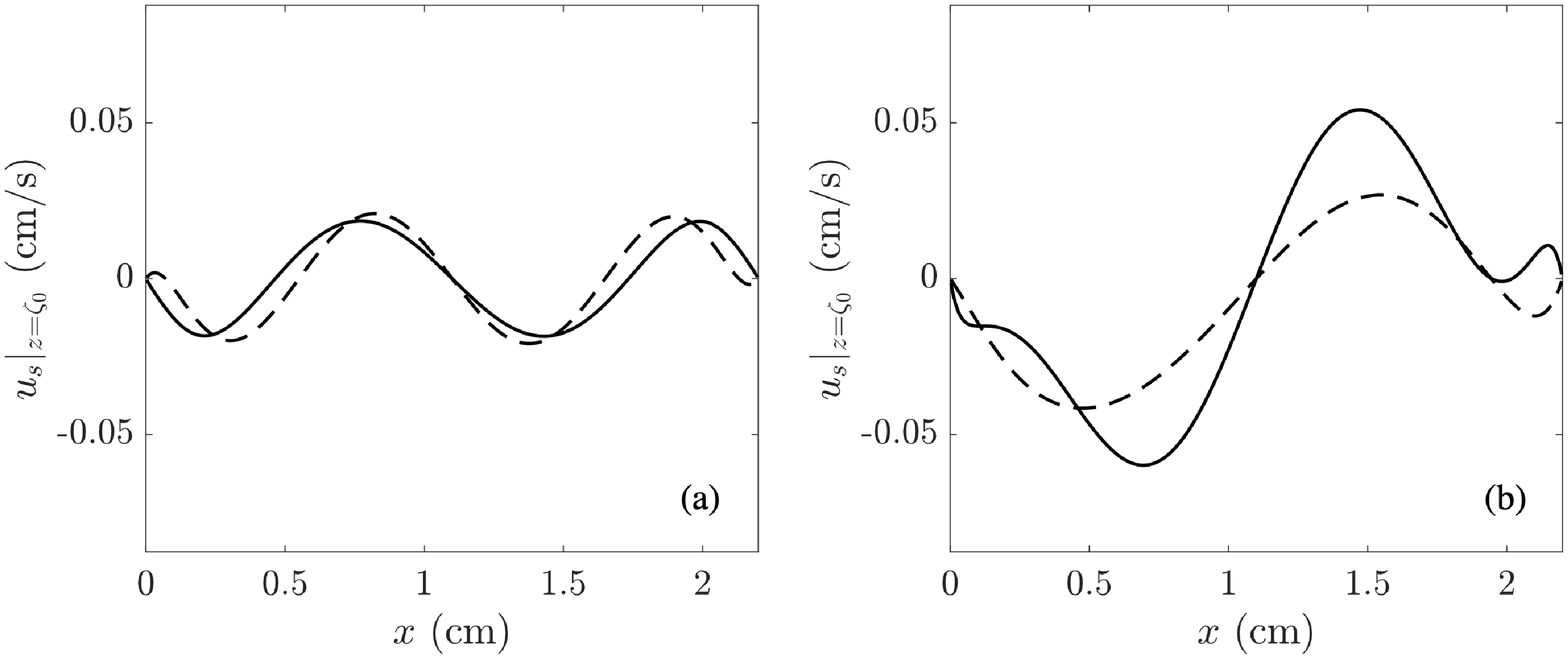}}
  \caption{The streaming flow velocity along the free surface when $a_{0}=0.2g_{0}$ and $f=8~\rm{Hz}$ for the (a) symmetric and (b) antisymmetric cases.
  The solid line is the result \eqref{streaming_velocity_2} truncated at $m, n = 200$ while the dashed line contains contributions from the gravest mode only.}
  \label{figure8}
\end{figure} 

The horizontal velocity along the meniscus, $u_{s}|_{z=\zeta_{0}}$, is shown in figure \ref{figure8}.
In both cases, the gravest mode gives the largest contribution to the streaming velocity.
With the contribution from the higher order waves, the magnitude decreases in the symmetric case and increases in the antisymmetric case.
The streaming velocity field is completely antisymmetric in the symmetric case, but lacks any definite symmetry in the antisymmetric case.
This asymmetry can be understood by considering the fact that the velocity of deep-water waves decreases exponentially from the mean water depth (i.e., from $z = 0$).
Since the steady liquid surface is higher on the hydrophilic side than on the hydrophobic side, the velocity induced by the meniscus wave is also higher on the hydrophilic side. 
From \eqref{streaming_velocity_1}, the wave-induced secondary flow inherits this strength distribution (see the schematic inset in figure \ref{figure11}), leading to an asymmetric streaming velocity distribution.

\subsubsection{Streaming near the lateral boundary}

Following \citet{batchelor2000introduction}, the streaming velocity for 2D flow near a flat vertical boundary is
\begin{align}
w_{s}|_{x=0,L}=-\frac{3}{4\omega}\sum_{n,m}\hat{w}_{n}\partial_{z}\hat{w}_{m}
\label{streaming_boundary}
\end{align}
where $\hat{w}_{n}$ is the vertical velocity amplitude of the $n^\text{th}$ mode at the lateral boundary. From \eqref{Solution_phi}, this is 
\begin{equation}
	\hat{w}_{n} = A_{n}k_{n}e^{k_{n}z}
		\begin{cases}
			1		&	x = 0, \\
			(-1)^n	&	x = L.
		\end{cases}
\end{equation}
Substituting this into \eqref{streaming_boundary} and applying linear extrapolation (see Appendix B), we find that
\begin{align}
w_{s}|_{x=0,L}=
\begin{cases}
-\frac{3}{4\omega}\sum\limits_{n,m}A_{n}A_{m}k_{n}k_{m}^{2}(1+zk_{n})(1+zk_{m}) & z\ge0, \\
-\frac{3}{4\omega}\sum\limits_{n,m}A_{n}A_{m}k_{n}k_{m}^{2}e^{(k_{n}+k_{m})z} & z<0.
\end{cases}
\label{lateral_streaming_velocity}
\end{align}
This expression is valid for both symmetric and anti-symmetric cases.

\begin{figure}
  \centerline{\includegraphics[scale=0.46]{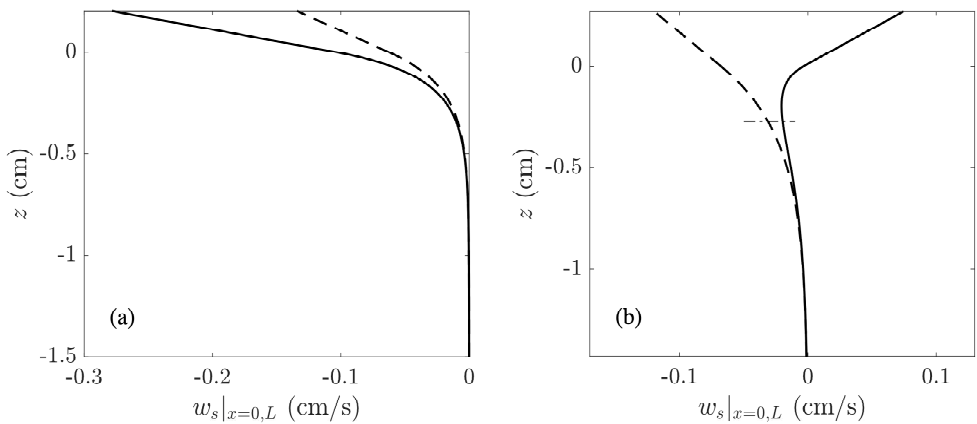}}
  \caption{The streaming flow velocity distribution along the lateral boundary when $a_{0}=0.2g_{0}$, $f=8~\rm{Hz}$ for the (a) symmetric and (b) antisymmetric cases. 
  The solid line is the result \eqref{streaming_boundary} with the sum terminated at $m,n = 200$ while the dashed line contains contributions from the gravest mode only.
  The upper limit of the plot is the position of the liquid surface at hydrophilic boundary.
  The dash-dotted line in (b) shows the liquid surface position at the hydrophobic boundary.}
  \label{figure9}
\end{figure} 

The vertical velocity distribution near the lateral boundary is shown in figure \ref{figure9}.
The contribution from the gravest mode is similar in both the symmetric and antisymmetric cases. 
The two cases become distinct once higher order contributions are included.
The streaming velocity becomes stronger, but remains monotonic and negative in the symmetric case.
In the antisymmetric case, the higher order contributions create a profile which is positive above $z = 0.013 cm$ and has a negative extremum about $0.2~\rm{cm}$ below the surface. 
This effect is due to phase modulation within the harmonic spectrum, $A_{n}$.
All the $A_{n}$ is positive in the symmetric case, so the components add constructively.
In the antisymmetric case, $A_{1}$ is negative and the other $A_{n}$ are positive, so interactions between the gravest and higher order modes interfere destructively with the contributions from the wave self-interactions.
Thus, in contrast to the monochromatic wave situation discussed by \citet{perinet2017streaming}, the lateral boundary streaming velocity profile is affected by the relative phases of the component waves.
In this experiment, the relative phases are controlled by the symmetry of static meniscus.
Note that the streaming velocity is left-right symmetric in the symmetric case, but asymmetric in the antisymmetric case because the free surface on the hydrophobic side is lower than that at hydrophilic side (see figure~\ref{figure9}b).

\subsubsection{Numerical simulation of streaming circulation}

Following \citet{perinet2017streaming}, we obtain the secondary flow in the bulk by using the steady streaming velocity profile as boundary conditions for the laminar flow solver in COMSOL Multiphysics\textsuperscript{\textregistered} using a fixed (i.e., not deforming) domain.
Taking the characteristic velocity $U$ to be the root mean square velocity of the surface streaming velocity distribution, and the characteristic length to be the cell width $L$, the maximum Reynolds number is $24$ for streaming circulation.
We simulate the steady secondary circulation at $7$ different oscillation amplitudes with the driving frequency $f=8~\rm{Hz}$.

\begin{figure}
  \centerline{\includegraphics[scale=0.55]{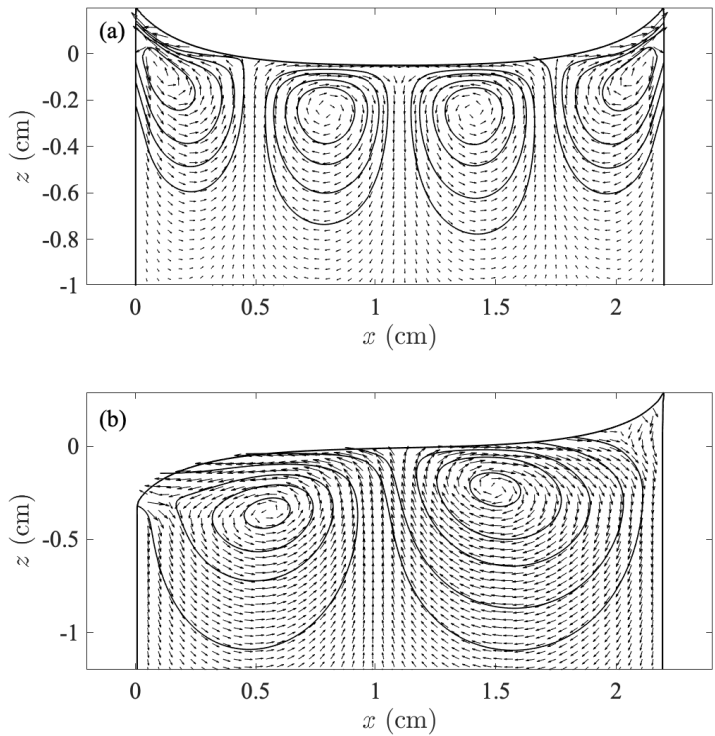}}
  \caption{The secondary flow velocity (areas) and streamfunction (contours) beneath the oscillating meniscus from numerical simulation for the (a) symmetric and (b) antisymmetric cases.
  The forcing amplitude is $a_{0}=0.2g_{0}$ and the driving frequency is $f=8~\rm{Hz}$.
  }
  \label{figure10}
\end{figure} 

The steady secondary circulation induced by the oscillation at $a_{0}=0.2g_{0}$ is shown in figure \ref{figure10}.
The numerical simulation successfully reproduces the four-vortex circulation structure in the symmetric case and dipole circulation structure in the antisymmetric case.
In the symmetric case, the generated circulation structure is nearly symmetric.
In the antisymmetric case, the circulation is deeper than in the symmetric case and stronger near the hydrophilic side---both in agreement with the experiments.

\begin{figure}
  \centerline{\includegraphics[scale=0.46]{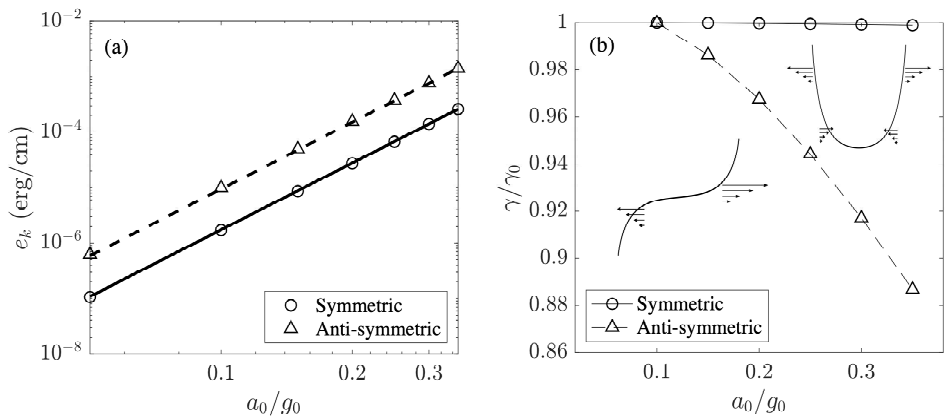}}
  \caption{Variation of (a) kinetic energy density, $e_k$, and (b) symmetry parameter, $\gamma/\gamma_0$, with $a_{0}/g_{0}$ from the numerical simulations of the symmetric and antisymmetric cases.
  The lines in (a) give the scaling $e_{k}\sim (a_{0}/g_{0})^{4}$.
  Note that the energy integration is performed over the same spatial region as in the experiments.
  The insets in panel (b) are two schematic plots illustrating the streaming velocity distribution under the free surface.}
  \label{figure11}
\end{figure} 

The numerical solutions follow the same scaling law, $e_k \sim (a_0/g_0)^4$, as the experiments, but the value of the energy is approximately one order of magnitude smaller (figure~\ref{figure11}a).
Since the secondary circulation is due to the quadratic nonlinearity, the energy of the secondary circulation scales with the amplitude of the meniscus wave to the fourth power.
Thus, small errors in the amplitude of the meniscus wave produce large errors in the energy of the secondary circulation.
Figure~\ref{fig:figure7} shows that the ratio of the amplitude of the meniscus wave derived from linear theory to the experimentally realized wave is approximately one-half in the symmetric case and two-thirds in the antisymmetric case.
These values are consistent with the observed errors in the energy density.
The other factor that can lead to such a deviation is that there exists a secondary flow convergence on $y$ direction at the observational plane near the hydrophobic angle (not shown in the paper). 
So, the convergence induced downwelling on the observational plane will also contribute to a portion of $e_{k}$.
 
As with the experiments, the numerical simulations retains a high degree of symmetry in the symmetric case, but looses symmetry as the forcing amplitude increases in the antisymmetric case (figure~\ref{figure11}).
However, the loss of symmetry is less dramatic in the numerical simulations than in the experiments.
This is also due to momentum injection caused by convergence of the flow in the $y$-direction, which will accelerates the 2D rotational motion on the hydrophilic side.

The numerical experiments were repeated with the surface streaming condition replaced by a free-slip boundary.
The four-vortex structure is lost in the symmetric case and the kinetic energy drops by an order of magnitude in both cases.
These results demonstrate that the free surface is a significant source of momentum for the secondary circulation and is important for determining its qualitative structure.

\section{Conclusion}

We observe the streaming circulation excited by standing capillary-gravity waves using PIV.
It is found that the structure of the streaming circulation is controlled by the static meniscus shape.
A semi-analytical theory is developed which explains the streaming pattern.

The experiment is designed to inhibit Faraday waves, so the streaming circulation is driven by resonant meniscus waves, rather than by Faraday waves as reported previously \citep{gordillo2014measurement,perinet2017streaming}.
Two different linear meniscus wave modes are generated by choosing the boundary wettability conditions.
In experiments with hydrophilic-hydrophilic boundary conditions, the static meniscus is symmetric and the gravest meniscus wave mode has two nodes.
In experiments with hydrophobic-hydrophilic boundary conditions, however, the static meniscus is antisymmetric and the gravest mode has a single node.
The experimentally measured free surface profiles are used to develop an analytical model for the linear meniscus waves.
This model reveals the meniscus waves are standing capillary-gravity waves with a spectrum determined by the Fourier spectrum of the static meniscus profile. 
The analytical results give the correct dominant mode number and structure, and predict the wave amplitude within an order of magnitude of the experimental results.

The secondary circulation induced by the meniscus wave is observed by broadcasting PIV images stroboscopically.
The secondary flow is originate from the Stokes boundary layer under the membrane formed by the cohesive PIV particles, where the Reynolds stress induced by the meniscus wave is balanced by the viscous stress.
The meniscus wave produces a four-vortex structure in the symmetric case, while a vortex dipole is generated in the antisymmetric case.
Regardless of the static surface profile, the total secondary circulation energy, $e_{k}$, scales like the forcing amplitude to the fourth power [i.e., $e_k \sim (a_{0}/g_{0})^{4}$].
In the symmetric case, the secondary flow retains a high degree of left-right symmetry as forcing amplitude is increased.
For the anti-symmetric case, the flow symmetry is broken, such an asymmetry comes from the exponential decay of the velocity from $z=0$, so the places with higher mean elevation has greater velocity under it. As a result, the anti-symmetric surface profile gives rise to the asymmetry of streaming circulation. 

Using the strategy suggested by \citep{perinet2017streaming}, we simulate the secondary flow by calculating the streaming velocities along the free surface and lateral boundaries and use these as boundary conditions for a laminar model. 
The simulated secondary flows have the same spatial structures, obey the same energy scaling, and have similar symmetry properties as the experimental flows.
The theoretical model reveals that the secondary flow is shaped by the entire wave spectrum rather than by a monochromatic wave, as has been studied extensively in previous literature.
In particular, the relative phases of harmonic modes significantly affect the resulting streaming velocity distribution.

Several limitations of this analysis should be reemphasized.
First, although the theory successfully captures the spatial structure of the meniscus wave, the solution is inaccurate in the region near the contact line because the we have assumed that the surface slope is small, whereas $\partial_{x}\zeta\sim 1$ near the contact line.
Second, the theory focuses on 2D dynamics since the experiment only measures a single cross-section of the flow field.
It is likely that 3D effects play some role in the secondary circulation, but data which could assess their significance is lacking.
For these reasons, the agreement between the theory and experiment is not fully quantitative.
In particular, the amplitudes of the meniscus waves and the strengths of the secondary circulations are underestimated.
Clearly, there remains much of interest to be further investigated.
	
	\section{Acknowledgement}
	We are grateful to Jie Yu, Robert Wilson and Dongping Wang for stimulating discussions. 
	We also thank Jun-Qiang Shi and Ze Chen for their participation in the early stage of the experiment. 
	This work was supported by the National Science Foundation of China through Grant No. 11572230 and 11772235, and a NSFC/RGC Joint Grant No. 11561161004.
	
	\section{Supplemental Materials}
	Supplementary movies that show the streaming circulations produced by oscillating meniscus (symmetric/antisymmetric, $a_0=0.2g_0$, $f=8~\rm{Hz}$) are available at ....
	
\appendix
\section{Faraday threshold}
This experiment focuses on streaming driven by a natural cavity mode excited through ordinary resonance, rather than a Faraday wave excited by parametric instability.
The experiment is therefore conducted in a stable zone of the Faraday instability.
Here, we demonstrate that the experiment is indeed within a stable zone.

The wavenumbers, $k_{lm}$, of linear modes in the working cell are
\begin{align}
	k_{lm}^{2}=\pi^2(\frac{l^2}{L_{x}^2}+\frac{m^2}{L_{y}^2}),
\end{align}
where $l$ and $m$ are the mode numbers in the $x$- and $y$-directions, respectively.
From the \citet{benjamin1954stability}, the mode amplitude $a_{lm}$ changes according to the differential equation
\begin{align}
	\frac{d^{2}a_{lm}}{dT^2}+[p_{lm}-2q_{lm}\cos(2T)]a_{lm}=0,
\end{align}
which is known as Mathieu's equation \citep{abramowitz1965handbook}. On this equation, $T=\frac{1}{2}\omega t$ is dimensionless time and the parameters $p_{lm}$ and $q_{lm}$ are
 \begin{align}
 	p_{lm}&=\frac{4k_{lm}\tanh(k_{lm}h)}{\omega^2}\left(g_{0}+\frac{\sigma}{\rho}k_{lm}^2\right),\\
 	q_{lm}&=2k_{lm}a_{0}\frac{\tanh(k_{lm}h)}{\omega^2}.
 \end{align}
The mathematical character of the Mathieu's equation is such that some modes grow exponentially in time
while others---in the stable zones---decay exponentially.
The decaying modes are not observable in this experiment, since their amplitudes rapidly become too small to measure.
For the forcing frequency $f=8~\rm{Hz}$ and amplitude $a_0 \le 0.36 g_0$, the parameters in Mathieu's equation are
\begin{align*}
	p_{10} &\approx 2.5,	&	p_{20} & \approx7.1, \\
	q_{10} &\le 0.39,		&	q_{20} &\le 0.79.
\end{align*}
For these values of the $p_{lm}$ the corresponding instability thresholds are $q^{T}_{10}\approx 2.6$ and $q^{T}_{20}\approx4.4$ \citep{abramowitz1965handbook}.
The values of $q_{lm}$ realized in the experiment are well below the Faraday instability threshold and the observed $(1,0)$ and $(2,0)$ modes are therefore ordinary resonant modes.

\section{Extrapolation method}

In this section, we describe the extrapolation method used to calculate the streaming velocity distribution along the meniscus at $z=\zeta_{0}(x)$.
Instead of extrapolating the secondary flow velocity directly, we extrapolate the primary flow (i.e., the linear meniscus wave) velocity to ensure continuity of the first-order velocity.
The primary flow velocities extrapolated to the static meniscus are
\begin{align}
u_{1}|_{z=\zeta_{0}}&\approx u_{1}|_{z=0}+\frac{\partial u_{1}}{\partial z}\Bigg|_{z=0}\zeta_{0}= -\cos \omega t\sum\limits_{n}^{\infty}A_{n}k_{n}\left[1+k_{n}\zeta_{0}\right]\sin k_{n}x,
\label{Extra_C1}\\
w_{1}|_{z=\zeta_{0}}&\approx w_{1}|_{z=0}+\frac{\partial w_{1}}{\partial z}\Bigg|_{z=0}\zeta_{0}=\phantom{-}\cos \omega t\sum\limits_{n}^{\infty}A_{n}k_{n}\left[1+k_{n}\zeta_{0}\right]\cos k_{n}x.
\label{Extra_C2}
\end{align}
The velocity tangent to the free surface is the projection of \eqref{Extra_C1} and \eqref{Extra_C2} onto the tangential direction $\hat{\bf{t}}$:
\begin{equation} \label{Extra_L}
\begin{aligned}
	u_{\parallel} &= (u_{1},w_{1})|_{z=\zeta_{0}}\cdot \hat{\bf{t}} \\
		&=-\cos\omega t\sum\limits_{n}^{\infty}A_{n}k_{n}\frac{1+k_{n}\zeta_{0}}{[1+(\partial_{x}\zeta_{0})^{2}]^{1/2}}
		\left(\sin k_{n}x	-\partial_{x}\zeta_{0}\cos k_{n}x\right)\\
&\approx -\cos \omega t\sum\limits_{n}^{\infty}A_{n}k_{n}(1+k_{n}\zeta_{0})\sin k_{n}x,
\end{aligned}
\end{equation}
where we have used the fact that $\partial_{x}\zeta_{0} \ll 1$ in the last step.
Again using the fact that $\partial_{x}\zeta_{0} \ll 1$, we estimate the tangential derivative of the tangential velocity as 
\begin{equation} \label{Extra_Lp}
	\partial_{\tau}u_{\parallel}
	\approx \partial _{x}u_{\parallel}
	\approx \cos \omega t\sum\limits_{n}^{\infty}A_{n} k_n(1+k_n\zeta_{0})\left(\partial_{xx}^{2}\zeta_{0}-k_{n}\right)\cos k_{n}x.   
\end{equation}
Substituting \eqref{Extra_L} and \eqref{Extra_Lp} into \eqref{streaming_velocity_1}, we find that the streaming velocity along the free surface is
\begin{equation} 
\begin{aligned}
	{\bf{u_{s}}} &=(u_{s},w_{s})
		=-\frac{3}{4\omega}\tilde{u}_{\parallel}\partial_{\tau}\tilde{u}_{\parallel}\hat{{\mathbf{t}}}\\
	&\approx-\frac{3}{4\omega}\hat{\textbf{\i}}\sum\limits_{n,m}^{\infty}A_{n}A_{m}k_{m} k_n (1+k_m\zeta_{0})(1+k_n\zeta_{0})\left(k_{n}-\partial_{xx}^{2}\zeta_{0}\right)\sin k_{m}x \cos k_{n}x,
\end{aligned}
\end{equation}
where we have used the facts that $\hat{\bf{t}}\cdot\hat{\textbf{\i}}\approx 1$, and $\hat{\bf{t}}\cdot\hat{\bf{k}} \approx 0$.
 
The process is similar for extrapolation on the lateral boundaries.
When $z\le0$, the streaming velocity is the same as for a flat surface.
When $z>0$, the vertical velocity at arbitrary position $z^{\prime}$ is estimated by 
\begin{equation}
	w_{1}|_{z=z^{\prime}}\approx w_{1}|_{z=0}+\frac{\partial w_{1}}{\partial z}|_{z=z^{\prime}}\zeta_{0}.
\end{equation}
 
It should be mentioned that the extrapolation for the velocity does not converge for arbitrary order of the Taylor expansion.
As $A_{n}$ converges like $n^{-5}$, the extrapolation polynomial in \eqref{Extra_C1} and \eqref{Extra_C2} should be no higher than the second order.

	\bibliographystyle{jfm}
	\bibliography{citation}

\end{document}